%% file: termgraph18.tex
\PassOptionsToPackage{dvipsnames}{xcolor}
\documentclass[copyright,creativecommons]{eptcs}
\providecommand{\event}{TERMGRAPH 2018} 

\usepackage{booktabs}   
\usepackage{subcaption} 
\input{preamble.tex}

\title{A Framework for Rewriting Families of String Diagrams}

\author{Vladimir Zamdzhiev
  \institute{Universit\'e de Lorraine, CNRS, Inria, LORIA, F 54000 Nancy, France}
}
\def\titlerunning{A Framework for Rewriting Families of String Diagrams}
\def\authorrunning{V. Zamdzhiev}

\begin{document}
\maketitle

\begin{abstract}
We describe a mathematical framework for
equational reasoning about infinite families of string diagrams which is
amenable to computer automation. The framework is based on context-free families
of string diagrams which we represent using context-free graph
grammars. We model equations between infinite families of diagrams using
rewrite rules between context-free grammars. Our framework represents
equational reasoning about concrete string diagrams and context-free families
of string diagrams using double-pushout rewriting on graphs and context-free
graph grammars respectively. We prove that our representation is sound by
showing that it respects the concrete semantics of string diagrammatic
reasoning and we show that our framework is appropriate for software
implementation by proving the membership problem is decidable.
\end{abstract}

\section{Introduction}
\begin{figure}
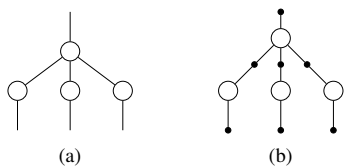

\cstikz[0.7]{string-individual.tikz}
\caption{\footnotesize{String diagram (a) and its string graph representation (b).}}\label{fig:string-diagram}
\end{figure}
String diagrams have found applications across a range of areas
in computer science and related fields such as concurrency
\cite{petri-nets}, systems theory~\cite{signal-flow}, quantum
computing~\cite{euler_necessity} and others.
A string diagram is a graph-like structure which consists of a collection of
\emph{nodes} together with a collection of (possibly open-ended) \emph{wires}
connecting nodes to each other (cf. Figure~\ref{fig:string-diagram}a).
\begin{figure}
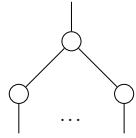

\centering
\cstikz[0.7]{string-family.tikz}
\caption{\footnotesize{Family of string diagrams.}}\label{fig:string-family}
\end{figure}
However, in some application scenarios, it is necessary to reason not just
about individual string diagrams, but about entire families of string diagrams (cf. Figure~\ref{fig:string-family}).
For example, in quantum computing, algorithms and protocols are often described
by a family of quantum circuits (diagrams), parameterised by the number of input qubits (wires).

However, as the size of a system grows, reasoning about large (families) of
string diagrams becomes cumbersome and error-prone. These issues can be
alleviated by using a diagrammatic proof assistant, such as
Quantomatic~\cite{quanto-cade}, which can automate the reasoning process.
Of course, this necessitates developing a formal framework which
can represent such families of string diagrams.

We will represent individual string diagrams using \emph{string
graphs}~\cite{kissinger_dphil}. A (directed) string graph is a (directed) graph with two
kinds of vertices -- \emph{wire} vertices and \emph{node} vertices. Wire
vertices have in-degree and out-degree at most one and are used to
represent the (open-ended) wires of string diagrams;
node vertices can be adjacent only to wire-vertices (cf. Figure~\ref{fig:string-diagram}b).

The primary contribution of this work is to improve the results of \cite{icgt}, which laid the foundation for
the representation of \emph{context-free families} of string diagrams and the methods used for
equationally reasoning about them. All of the results in this paper are described in detail in the author's PhD thesis \cite{vladimir-dphil}.
We will represent families of string diagrams and equational reasoning for them by context-free grammars of string graphs and DPO rewriting of these grammars, respectively.

\section{Background}

We begin by introducing some notation and, because of lack of space, briefly recalling the theory
of \emph{B-edNCE graph grammars} (see~\cite{c-ednce} or~\cite[Chapter
2]{vladimir-dphil} for the full definitions).
Throughout the rest of the paper, we consider graphs whose vertices
are labelled over an alphabet $\Sigma$ and whose edges are labelled over an
alphabet $\Gamma$.
$\Delta \subseteq \Sigma$ is the \emph{alphabet of terminal
vertex labels};
$\mathcal{N} 
\subseteq \Delta$ is the \emph{alphabet of node-vertex
labels} and $\mathcal{W} = \Delta - \mathcal{N}$ is the alphabet of
\emph{wire-vertex labels}.

\begin{definition}[Graph \cite{c-ednce}]\label{def:graph}
A \emph{graph} over an alphabet of vertex labels $\Sigma$ and an alphabet of
edge labels $\Gamma$ is a tuple $H = (V, E, \lambda)$, where $V$ is a finite
set of nodes, $E \subseteq \{(v, \gamma, w) | v, w \in V, v \not= w, \gamma \in
\Gamma\}$ is the set of edges and $\lambda : V \to \Sigma$ is the vertex
labelling function.
\end{definition}

\begin{remark}
The framework which we describe works with both directed and undirected graphs,
where the latter requires only a small simplification of some definitions. To
retain generality, all definitions within the paper are stated for directed
graphs, but many of our examples show undirected graphs (as they make for more interesting examples).
In our notion of graph, self-loops are not allowed and parallel edges are allowed as long as they have
distinct labels.
This requirement is commonly imposed by node-replacement graph grammars in the litereature.
\end{remark}

\begin{definition}[Extended Graph \cite{c-ednce}]
\label{def:extended-graph}
  An \textit{Extended Graph} over $\Sigma$ and $\Gamma$ is a pair $(H,
  C),$ where $H$ is a graph and $C \subseteq \Sigma
  \times \Gamma \times \Gamma \times V_H \times \{in, out\}$. $C$ is called
  a \textit{connection relation} and its elements $(\sigma, \beta, \gamma, x,
  d)$ are called \textit{connection instructions}.
  The set of all extended graphs over $\Sigma$ and $\Gamma$ is denoted
  by $EGR_{\Sigma, \Gamma}$.
\end{definition}
An
\emph{extended graph} provides the necessary information on how a
specific graph can be used to replace a nonterminal vertex and connect it to
the local neighbourhood of the nonterminal vertex that is to be replaced.
In the literature, extended graphs are commonly referred to as \emph{graphs
with embedding}~\cite{c-ednce}.

If we are given a mother graph $(H, C_H)$ with a nonterminal vertex
$v \in H$ and a daughter graph $(D, C_D)$, then
the \textit{substitution} of $(D, C_D)$ for $v$ in $(H, C_H)$, denoted by
$(H, C_H)[v/(D,C_D)]$, is given by the extended graph
constructed in the following way:
for every connection instruction $(\sigma, \beta, \gamma,x,in) \in C_D$
and for every $\sigma$-labelled vertex $w$ in the
mother graph for which there is a $\beta$-labelled edge going \textbf{in}to the
nonterminal vertex $v$ of the mother graph, then the
substitution process will establish a $\gamma$-labelled edge from $w$ to $x$.
This
should become more clear after referring to Example \ref{ex:derivation}.
The meaning for $(\sigma, \beta,
\gamma,x,out)$ is analogous.
Next, we define the concept of an edNCE Graph Grammar. edNCE is an
abbreviation for \textbf{N}eighbourhood \textbf{C}ontrolled \textbf{E}mbedding
for \textbf{d}irected graphs with dynamic \textbf{e}dge relabelling.

\begin{definition}[edNCE Graph Grammar \cite{c-ednce}]
  An \textit{edNCE Graph Grammar} is given by a pair $G = (P,S)$, where
    $P$ is a finite set of productions and
    $S \in \Sigma - \Delta$ is the initial nonterminal label.
  Productions are of the form $X \rightarrow (D, C)$, where $X \in \Sigma -
  \Delta$ is a nonterminal label and $(D, C) \in EGR_{\Sigma, \Gamma}$ is an
  extended graph.
For a production $p:= X \to (D,C)$, we shall say that the \emph{left-hand side}
of $p$ is $X$ and denote it with $lhs(p)$. The \emph{right-hand side} of $p$ is
the extended graph $(D,C)$ and we denote it with $rhs(p)$. Vertices
which have a label from $\Delta$ are called \emph{terminal vertices} and
vertices with labels from $\Sigma - \Delta$ are called \emph{nonterminal
vertices}. An
(extended) graph is called terminal if all of its vertices are
terminal.
\end{definition}

Instead of presenting grammars using set-theoretic notation, we will often
present them graphically as it is more compact and intuitive. We will use the
same notation as in \cite{c-ednce}, which we now explain.
The grammar of Figure~\ref{fig:derivation} has three productions.
In each production, the nonterminal symbol
is written in the top-left corner of the box. The content of the box
is simply the graph which replaces its corresponding nonterminal in a derivation.
Any edges crossing the box are the connection instructions which indicate how
to connect the graph within the box to its outside context.

\begin{definition}[Extended Graph homomorphism]\label{def:extended-graph-homomorphism}
Given two extended graphs $(H,C_H), (K,C_K) \in EGR_{\Sigma,\Gamma}$, an
\emph{extended graph homomorphism} between $(H, C_H)$ and $(K,C_K)$ is a
function $f: V_H \to V_K$, such that $f$ is a graph homomorphism from
$H$ to $K$ and if  $(\sigma, \beta, \gamma, x, d) \in C_H$ then
$(\sigma, \beta, \gamma, f(x), d) \in C_K$.
\end{definition}

\begin{definition}[Derivation \cite{c-ednce}]
For a graph grammar $G=(P,S)$ and extended graphs
$H, H',$ let $v \in V_H$ be a non-terminal vertex with label $X$
and $p: X\to(D,C)$ be a production (copy) of the
grammar, such that $H$ and $D$ are disjoint. We say $H \Longrightarrow_{v,p}
H'$ is
a \textit{derivation step} if $H'=H[v/(D,C)]$. A
sequence of derivation steps $H_0 \Longrightarrow_{v_1,p_1} H_1
\Longrightarrow_{v_2,p_2} \cdots \Longrightarrow_{v_n,p_n} H_n$ is called a
\textit{derivation}. We write $H \Longrightarrow_* H'$ if there
exists a derivation from $H$ to $H'$. A derivation $H \Longrightarrow_*
H'$ is \emph{concrete}
if $H'$ is terminal.
\end{definition}

\begin{definition}[Graph Grammar Language \cite{c-ednce}]
A \emph{sentential form} of an edNCE grammar
$G=(P,S)$ is a graph $H$ such that
$sn(S,z) \Longrightarrow_* H$ for some $z$,
where $sn(S,z)$ is the (extended) graph
that has only one vertex given by $z$, its label is $S$ and the graph has no edges 
and no connection instructions.
The graph language induced by $G$
is the set of all terminal sentential forms modulo graph isomorphism.
\end{definition}

\begin{definition}[B-edNCE grammar \cite{c-ednce}]
\label{def:b-ednce}
An edNCE grammar $G=(P,S)$ is
\emph{boundary}, or a B-edNCE grammar, if for every production
$X \to (D,C)$, we have that
$D$ contains no edges between nonterminal vertices and
$C$ does not contain connection instructions of the form
$(\sigma, \beta, \gamma, x, d)$ where $\sigma$ is a nonterminal label.
\end{definition}

\begin{example}\label{ex:derivation}
The language of the B-edNCE grammar from Figure~\ref{fig:derivation}(a) consists of the string graph
representations of the string diagrams from Figure~\ref{fig:string-family}.
The (concrete) derivation which produces the string graph from Figure~\ref{fig:string-diagram}(b) is shown
in Figure~\ref{fig:derivation}(b).
\begin{figure}[h]
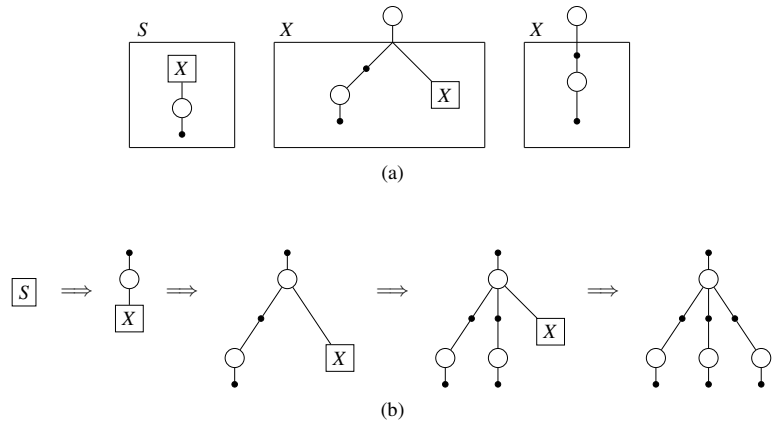

\centering
\stikz[0.7]{gg_kn_grammar.tikz}
\caption{A B-edNCE grammar of undirected graphs (a) and a concrete derivation from it (b).}\label{fig:derivation}
\end{figure}
\end{example}

\section{Reasoning about families of string diagrams}
\begin{figure}
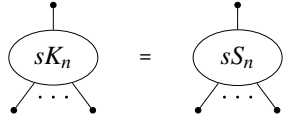

  \cstikz[0.7]{family-rewrite.tikz}
  \caption{Equational schema}\label{fig:family-rewrite}
\end{figure}
In this section, we present our framework which allows us to equationally reason
about entire families of string diagrams (as opposed to individual string diagrams). As
a motivating example, we consider the local complementation
rule~\cite{euler_necessity} of the ZX-calculus (used in quantum computing), which is crucial for
establishing a decision procedure for equality of diagrams in the calculus.
The essential data of the rule is given by the equational \emph{schema} in Figure~\ref{fig:family-rewrite}.
where $sK_n$ is the complete \emph{string} graph on $n$ vertices (consisting of
$n$ nodes connected to each other via wires consisting of 1 wire vertex)
and $sS_n$ is the star \emph{string} graph on $n$ vertices (the string graph representation of Figure~\ref{fig:string-family}).
In both cases, every node has also exactly one open-ended wire attached to it.
\subsection{Context-free families of string graphs}
As we saw in Example~\ref{ex:derivation}, edNCE grammars may represent the $\{sS_n\}_{n \in \mathbb N}$ family, but unfortunately,
they cannot represent the $\{sK_n\}_{n \in \mathbb N}$ family~\cite[Section 3.3]{vladimir-dphil}. However, it is well-known that (B)-edNCE grammars \emph{can} represent the family of complete graphs and
if we think of the edges of a complete graph as if they are representing a wire
with a single wire vertex, then we can recover the latter family. This idea
may be easily formalized and allows us to slightly extend the expressive
power of edNCE grammars, while still retaining crucial decidability properties (like the membership problem).

As a first step, we refine the alphabet of edge labels by introducing
$\mathcal E \subseteq \Gamma$ to be the alphabet of \emph{encoding}
edge labels. Essentially, the idea described above is formalized by using graph
grammars which generate graphs that contain some encoding edges which are
subsequently decoded using a simple confluent and terminating DPO rewrite
system.
\begin{definition}[Encoded string graph \cite{icgt}]
An \textit{encoded string graph} is a string graph where we additionally allow
edges with labels $\alpha \in \mathcal E$ to connect pairs of node-vertices.
Edges labelled by some $\alpha \in \mathcal E$ will be called \emph{encoding}
edges.
\end{definition}

\begin{definition}[Decoding system \cite{icgt}]
A \textit{decoding system} $T$ is a set of DPO rewrite rules of the form:
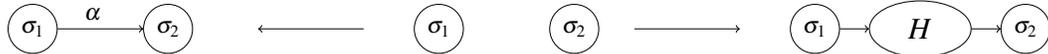
\cstikz[0.8]{simple-dpo-def.tikz}
one for every triple $(\alpha, \sigma_1, \sigma_2) \in \mathcal E \times
\mathcal N \times \mathcal N,$
where the LHS consists of a single edge with encoding label $\alpha \in
\mathcal E$ connecting a $\sigma_1$-labelled node-vertex to a
$\sigma_2$-labelled node-vertex, and the RHS is a string graph which contains
the same two node-vertices and at least one additional vertex while containing
no inputs, outputs, or encoding labels.
\end{definition}

\begin{theorem}
Any decoding system $T$ is confluent and terminating.
\end{theorem}

Given an (encoded string) graph, \emph{decoding} is the process of applying all
of the rules of $T$ to the graph. As the above theorem shows, this is a
very simple process which may even be done in a single step.
If $H$ is an encoded string graph, we shall say that $H'$ has been
\emph{decoded} from $H$, and denote this with $H \Longrightarrow_*^T H',$ if
the graph $H'$ is the result of applying all rules from $T$ to $H$, such that
$H'$ contains no encoding edges.

\begin{lemma}\label{lem:decoding}
Given two graphs $H, H'$ with $H \Longrightarrow_*^T H'$, where $T$ is
a decoding system, then $H$ is an encoded string graph iff $H'$ is a string
graph.
\end{lemma}

\begin{definition}[Encoded B-edNCE grammar]\label{def:encoded-b-ednce}
An \emph{encoded B-edNCE grammar} is a pair $B=(G,T)$, where $G$ is a B-edNCE
grammar and $T$ is a decoding system.
A \textit{concrete derivation} for an encoded B-edNCE grammar $B = (G,T)$ with
$S$ the initial nonterminal, consists of a concrete derivation in $G$
followed by a decoding in $T$, which we denote as
$sn(S,z) \Longrightarrow_*^G H_1 \Longrightarrow_*^T
H_2$ or
simply with $sn(S,z) \Longrightarrow_*^B H_2$ if the graph $H_1$ is not
relevant for
the context.
\end{definition}

\begin{definition}[B-ESG grammar]\label{def:besg} \rm
A \emph{Boundary Encoded String Graph} (\textit{B-ESG}) grammar is an encoded B-edNCE grammar $B = (G, T),$
subject to some additional coherence conditions, which we omit for lack
of space (see~\cite[Chapter 4]{vladimir-dphil}).
\end{definition}

The coherence conditions from the above definition are both necessary (up to normal form) and sufficient to generate
languages of string graphs, as the next two theorems show.

\begin{theorem}\label{thm:besg_language}
  Every graph in the language of a B-ESG grammar is a string graph.
\end{theorem}

\begin{theorem}\label{thm:necessary}
Given an encoded B-edNCE grammar $B=(G,T),$ such that $L(B)$ is a language
consisting of string graphs, then there exists a B-ESG grammar $B'=(G',T)$,
such that $L(B') = L(B).$ Moreover, $G'$
can be constructed effectively from $G$ and $L(G') = L(G)$.
\end{theorem}

\begin{example}\label{ex:complete}
The language of the B-ESG grammar $B=(G,T)$ below is the family $\{sK_n\}_{n \in \mathbb N}.$
\cstikz[0.7]{complete-string-graphs.tikz}
$sK_4$ is derived by first generating a graph with encoding edges (labelled $\alpha$) followed by decoding:
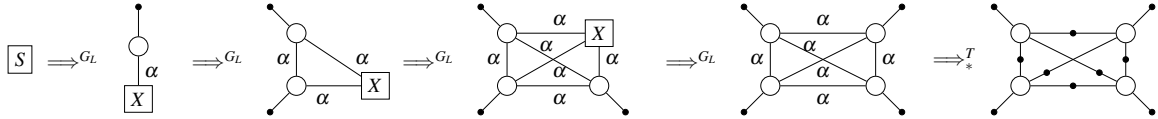
\cstikz[0.7]{sk3-derive.tikz}
The grammar from Example~\ref{ex:derivation} is also a B-ESG grammar when
equipped with any decoding system (which does not have an effect on the
generated language, because the grammar does not contain encoding edges).
\end{example}

\begin{theorem}\label{thm:member} \rm
  The membership problem for B-ESG grammars is decidable.
\end{theorem}

\begin{definition}[Grammar homomorphism]\label{def:grammar-homomorphism}
Given two edNCE grammars
$G_1 = (P_1, S_1)$ and
$G_2 = (P_2, S_2)$, a \emph{grammar
homomorphism} from $G_1$ to $G_2$ is a function $m : P_1 \to P_2$, together
with a collection of extended graph homomorphisms (cf. Definition~\ref{def:extended-graph-homomorphism}) $m_{p_i} : rhs(p_i) \to
rhs(m(p_i))$ one for each production $p_i \in P_1$, such that $lhs(p_i) =
lhs(m(p_i)).$
\end{definition}

\begin{definition}[Category of B-ESG grammars]
The category of B-ESG grammars over a decoding system $T$, denoted
\textbf{B-ESG}$_T$, or simply \textbf{B-ESG} if $T$ is clear from the context,
has objects B-ESG grammars $B = (G,T)$. A morphism $h$ between two B-ESG
grammars $B_1 = (G_1, T)$ and $B_2 = (G_2, T)$ is a grammar
homomorphism $h : G_1 \to G_2$.
\end{definition}

\begin{theorem}
\textbf{B-ESG}$_T$ is a \emph{partially adhesive} category (cf. \cite{kissinger_dphil}).
\end{theorem}

The above theorem means that DPO rewriting of
B-ESG grammars \emph{themselves} is well-behaved, provided that some additional matching
conditions (which are fully characterised) are satisfied~\cite[Chapter
5]{vladimir-dphil}.

\subsection{Rewrite schemas for families of string graphs}

We saw how to represent families of string graphs, next we explain how to
represent equational schemas between such families. That is, we wish to
establish a constructive bijection between the graphs of one family and the
graphs of the other. We may do so if we require that the pair of B-ESG grammars
have a 1-1 correspondence between their productions and so do the nonterminal
vertices in corresponding productions. This allows us to perform
\emph{parallel} concrete derivations between the two grammars and we may thus
establish the constructive bijection we required.

\begin{definition}[Extended graph rewrite rule]
An \emph{extended graph rewrite rule} is a pair of
monomorphisms $L \xleftarrow{l} I \xrightarrow{r} R,$ where all objects
are extended graphs.
\end{definition}

\begin{definition}\label{def:subst_embed}
Given extended graphs $G$ and $D$ where $x \in V_G$ is a (nonterminal) vertex,
and given monomorphisms $m_1 : G \to G'$ and $m_2 : D \to D',$ then
\[ m_3(v) := \begin{cases} m_1(v) & \text{ if } v \in V_G\\
                        m_2(v) & \text{ if } v \in V_D
          \end{cases}
\]
is the induced \emph{substituted monomorphism} $m_3 : G[x/D] \to G'[m_1(x)/D'],$ which we
denote by $SM(m_1,m_2,x)$.
\end{definition}

\begin{definition}[Rewrite rule substitution]
Given extended graph rewrite rules $B_1 := L_1 \xleftarrow{l_1} I_1
\xrightarrow{r_1} R_1$ and $B_2 := L_2 \xleftarrow{l_2} I_2 \xrightarrow{r_2}
R_2$, with (non-terminal) vertex $v \in I_1$ then the \emph{substitution} of $B_2$ for
$v$ in $B_1$, denoted $B_1[v/B_2]$ is given by the extended graph rewrite rule
$B_3 := L_3 \xleftarrow{l_3} I_3 \xrightarrow{r_3} R_3$, where
$L_3 := L_1[l_1(v)/L_2],$
$I_3 := I_1[v/I_2],$
$R_3 := R_1[r_1(v)/R_2],$
$l_3 :=SM(l_1, l_2, v)$ and $r_3 := SM(r_1, r_2, v)$. 
\end{definition}

\begin{definition}[B-edNCE Pattern]
A B-edNCE \emph{pattern} is a triple of B-edNCE grammars $B:= G_L
\xleftarrow l G_I \xrightarrow r G_R$,
where $l$ and $r$ are grammar monomorphisms which are
bijections between the productions of all three grammars. Moreover,
$l$ and $r$ are also label-preserving bijections between the nonterminals in
corresponding productions of the grammars. In addition, all three grammars
have the same initial nonterminal label. If $p$ is a production in $G_L$,
then $B_p$ will refer to the extended graph rewrite rule $rhs(p)
\xleftarrow{l_{p'}} rhs(p') \xrightarrow{r_{p'}} rhs(p''),$
where $p'$ and $p''$ are the corresponding productions of $p$ in $G_I$ and
$G_R$ respectively,
and $l_{p'}$ and $r_{p'}$ are the components (cf. Definition~\ref{def:grammar-homomorphism}) of the monomorphisms $l$ and $r$ at production $p'$.
\end{definition}

These conditions ensure that we may perform parallel derivations in the sense that at each step we apply
corresponding productions and replace corresponding nonterminals in all three grammars.

\begin{definition}[B-edNCE Pattern Instantiation]
Given a B-edNCE pattern $G_L \xleftarrow{l} G_I \xrightarrow{r} G_R$, a
parallel instantiation is a
triple of concrete derivation sequences of the following form:
\begin{align*}
sn(S, v_1) &\Longrightarrow_{v_1,l(p_1)}^{G_L} H_1'
\Longrightarrow_{l(v_2),l(p_2)}^{G_L} H_2'
\Longrightarrow_{l(v_3),l(p_3)}^{G_L} \cdots
\Longrightarrow_{l(v_n),l(p_n)}^{G_L} H_n'\\
sn(S, v_1) &\Longrightarrow_{v_1,p_1}^{G_I} H_1
\Longrightarrow_{v_2,p_2}^{G_I} H_2 \Longrightarrow_{v_3,p_3}^{G_I} \cdots
\Longrightarrow_{v_n,p_n}^{G_I} H_n\\
sn(S, v_1) &\Longrightarrow_{v_1,r(p_1)}^{G_R} H_1''
\Longrightarrow_{r(v_2),r(p_2)}^{G_R} H_2''
\Longrightarrow_{r(v_3),r(p_3)}^{G_R} \cdots
\Longrightarrow_{r(v_n),r(p_n)}^{G_R} H_n''
\end{align*}
The language
of $B,$ denoted $L(B),$ is the set of all graph rewrite rules
$H_n' \xleftarrow{l_n} H_n \xrightarrow{r_n} H_n''$
obtained by
performing concrete parallel derivations, where $l_n$ and $r_n$ are
induced by the derivation process.
\end{definition}

\begin{definition}[Production input/output/isolated vertex]
Given a B-ESG grammar $B$, we say that a wire-vertex $w$ is a \emph{production
input (output)} if its in-degree (out-degree) is zero and it has no
incoming (outgoing) connection instructions. $w$ is a \emph{production
isolated wire-vertex} if it is both a production input and a production output.
\end{definition}

\begin{definition}[B-ESG rewrite rule]
A \emph{B-ESG rewrite rule} is a span of monos $B_L \xleftarrow l B_I
\xrightarrow r B_R,$ where 
$B_L = (G_L,T),
B_I = (G_I, T), B_R = (G_R, T),$ such that $G_L \xleftarrow l G_I \xrightarrow r
G_R$ is a B-edNCE pattern such that for every triple of corresponding
productions $p_L, p_I, p_R$ in $G_L, G_I, G_R$ respectively, we have:
\begin{description}
\item[Boundary:] $p_I$ contains only nonterminal vertices and isolated
wire-vertices and it contains no edges, connection instructions or
node-vertices.
\item[IO1:] $l$ and $r$ are surjections on the production inputs (outputs)
between $p_I$ and $p_L$, $p_I$ and $p_R$ respectively.
\item[IO2:] For every wire-vertex $w \in p_I$, $l(w)$ and $r(w)$ are both a
production input (output) in $p_L$ and $p_R$ respectively.
\end{description}
Moreover, the grammars $G_I, G_L, G_R$ must be in a certain normal form (cf. ~\cite[Section 5.3]{vladimir-dphil}).
\end{definition}

\begin{definition}[B-ESG Rewrite Rule Instantiation]\label{def:b-esg-inst}
Given a B-ESG rewrite rule $B:= B_L \xleftarrow{l} B_I \xrightarrow{r} B_R$, a
parallel instantiation is a B-edNCE pattern instantiation for
$G_L \xleftarrow{l} G_I \xrightarrow{r} G_R$ followed by a decoding:
\begin{align*}
sn(S, v_1) &\Longrightarrow_{v_1,l(p_1)}^{G_L} H_1'
\Longrightarrow_{l(v_2),l(p_2)}^{G_L} H_2'
\Longrightarrow_{l(v_3),l(p_3)}^{G_L} \cdots
\Longrightarrow_{l(v_n),l(p_n)}^{G_L} H_n'
\Longrightarrow_*^T F'\\
sn(S, v_1) &\Longrightarrow_{v_1,p_1}^{G_I} H_1
\Longrightarrow_{v_2,p_2}^{G_I} H_2 \Longrightarrow_{v_3,p_3}^{G_I} \cdots
\Longrightarrow_{v_n,p_n}^{G_I} H_n
\Longrightarrow_*^T F\\
sn(S, v_1) &\Longrightarrow_{v_1,r(p_1)}^{G_R} H_1''
\Longrightarrow_{r(v_2),r(p_2)}^{G_R} H_2''
\Longrightarrow_{r(v_3),r(p_3)}^{G_R} \cdots
\Longrightarrow_{r(v_n),r(p_n)}^{G_R} H_n''
\Longrightarrow_*^T F''
\end{align*}
The language of $B,$ denoted $L(B),$ is the set of all rewrite rules $F'
\xleftarrow{l_F} F \xrightarrow{r_F} F''$ obtained by performing
parallel derivations, where the embeddings
$F' \xleftarrow{l_F} F \xrightarrow{r_F} F''$ are induced by the derivation process.
\end{definition}

\begin{theorem}\label{thm:b-esg-inst}
The language of every B-ESG rewrite rule consists solely of string graph
rewrite rules.
\end{theorem}

\begin{example}\label{ex:besg-rewrite-rule}
Let $B_L=(G_L,T)$ be the B-ESG grammar representing $\{sK_n\}_{n \in \mathbb N}$ from Example~\ref{ex:complete}
and let $B_R=(G_R,T)$, where $G_R$ is the grammar representing $\{sS_n\}_{n \in \mathbb N}$ from Figure\ref{fig:derivation}.
Given this data, there is a unique (and constructive) choice for a B-ESG grammar $B_I$ and embeddings $l,r$, such that
$B:= B_L \xleftarrow{l} B_I \xrightarrow{r} B_R$ is a B-ESG rewrite rule. $B_I = (G_I,T),$ where $G_I$ is given by:
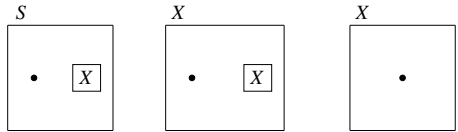
\cstikz[0.7]{interface.tikz}
and $l$ and $r$ are the obvious grammar embeddings.
A derivation of a string graph rewrite rule relating $sK_4$ and $sS_4$ is given by a derivation involving 4 (parallel) steps in the B-edNCE grammars and a decoding:
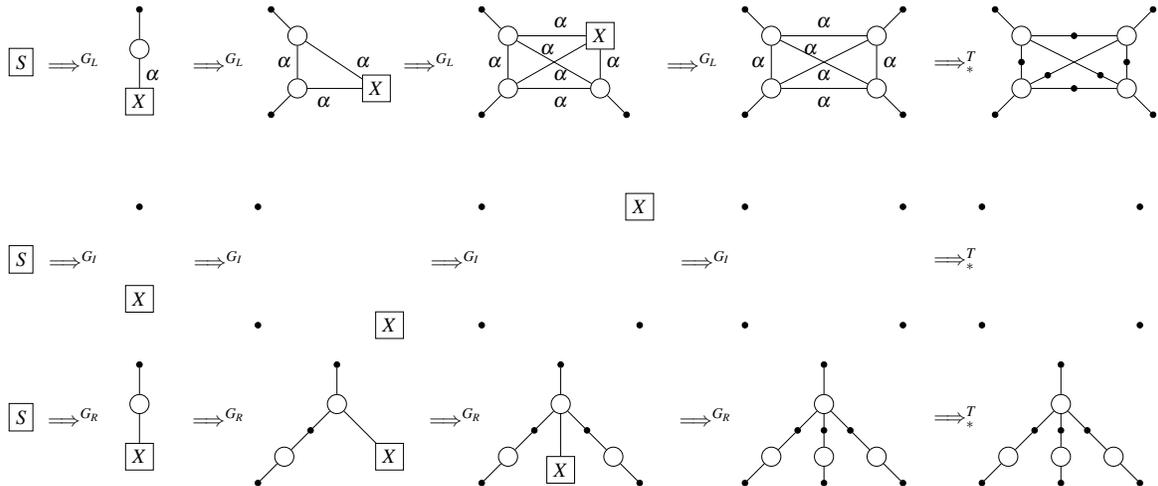
\cstikz[0.7]{pattern-derive2.tikz}
where there are obvious induced embeddings of the middle string graph into the other two (details omitted).
\end{example}

\subsection{B-ESG grammar rewriting}
Next, we show how to model equational reasoning by
applying an equational schema to a context-free family of diagrams.
We begin by
first showing that DPO rewriting (which we use to model equational reasoning for graphs and grammars)
behaves well with respect to graph substitution (which we use for language
generation).
We use the notation $H \leadsto^m_B H'$ to indicate that the DPO rewrite of the
(extended) graph $H$ using the (extended) graph rewrite rule $B$ at matching
$m$ is the (extended) graph $H'$.

\begin{theorem}\label{thm:subst_rewrite}
Given boundary extended graphs $H, H', D, D',$ such that $H
\leadsto_{B_1}^{m_1} H'$ and $D \leadsto_{B_2}^{m_2} D',$ where $B_1 := L_1
\xleftarrow{l_1} I_1 \xrightarrow{r_1} R_1$, $B_2 := L_2
\xleftarrow{l_2} I_2 \xrightarrow{r_2} R_2,$ $v \in V_{I_1}$
and where $m_1, m_2$ are matchings (subject to some additional conditions),
then $H[m_1 \circ l_1(v)/D] \leadsto_{B_3}^{m_3} H'[f_1\circ r_1(v)/D'],$ where
$B_3 := B_1[v/B_2]$ and $m_3 = SM(m_1, m_2, l_1(v))$. In terms of diagrams,
given the following two DPO rewrites:
\cstikz{dpo.tikz}
then the following diagram is also a DPO rewrite:
\cstikz{dpo3.tikz}
where each $x_3$ is the obvious substituted monomorphism (cf. Definition~\ref{def:subst_embed}).
\end{theorem}

\begin{definition}[B-ESG rewrite]\label{def:b-esg-rewrite}
Given a B-ESG rewrite rule $B= B_L \xleftarrow l B_I \xrightarrow r B_R$
with initial nonterminal label $S$
and a B-ESG grammar $B_H,$ such that $B_H$ is in normal form (cf.~\cite[Section 5.3]{vladimir-dphil}),
then we will say that the
\emph{B-ESG rewrite} of $B_H$ using $B$ over a matching $m: G_L \to
G_H$, is the B-ESG
grammar $B_M
= (G_M, T),$ denoted by $B_H \leadsto_{B,m} B_M$, where $G_M$ is given by
the DPO rewrite:
\cstikz{b-esg-dpo.tikz}
\end{definition}

\begin{theorem}\label{thm:final}
Given a B-ESG rewrite $B_H \leadsto_{B,m} B_M$, as in
Definition~\ref{def:b-esg-rewrite}, where the matching $m$ satisfies some additional conditions (cf.~\cite[Section 5.4]{vladimir-dphil}), then the rewrite $B_H \leadsto_{B,m} B_M$ is admissible
with respect to $L(B)$ in the following sense: if $(K,K')$ is a parallel instantiation (cf. Definition~\ref{def:b-esg-inst})
of $(B_H, B_M)$, then there exists a sequence of string graph rewrite rules $s_1, \ldots, s_n \in L(B),$
such that $K \leadsto_{s_1} \cdots \leadsto_{s_n} K'.$
\end{theorem}

\begin{example}
Consider the equational schema in Figure~\ref{fig:final-schema}.
It can be derived from its left-hand side by applying the equational
schema from Figure~\ref{fig:family-rewrite}. Moreover, this application respects
the concrete semantics of the graph families in the sense that every
\emph{instantiation} of the former schema can be obtained by applying
a specific \emph{instantiation} of the latter schema.
\begin{figure}
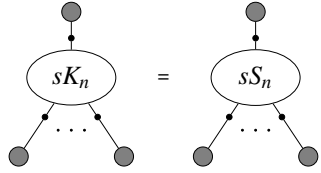

\cstikz[0.7]{final-schema.tikz}
\caption{Derived equational schema.}\label{fig:final-schema}
\end{figure}
We now show how to model this using our framework.
The left-hand side of Figure~\ref{fig:final-schema} is represented by the B-ESG grammar
$B_H = (G_H,T)$ given below (left). By performing a B-ESG rewrite using the
B-ESG rewrite rule from Example~\ref{ex:besg-rewrite-rule}, we get the grammar
$B_P=(G_P,T)$ given below (right), which correctly represents the right-hand
side of Figure~\ref{fig:final-schema}.
\cstikz[0.67]{lhs.tikz}
Moreover, this rewrite respects the concrete semantics, because for any
parallel instantiation $(K, K')$ of the two grammars above, the string graph $K$ may be
rewritten into the string graph $K'$ by applying a concrete string graph rewrite rule induced by the
B-ESG rewrite rule from Example~\ref{ex:besg-rewrite-rule}.
\end{example}

\section{Use case: generalised bialgebra rule}

In the previous section we presented a running example which showed how to
model the local complementation rule for the ZX-calculus. We now provide
another example, namely of the Generalised Bialgebra rule, which is an
important distributivity law for several diagrammatic calculi, including the
ZX-calculus. We will show how this can be represented using our framework.

The generalised bialgebra rule is given by the equational schema of string diagrams given by (1):
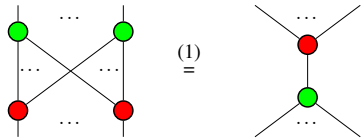
\cstikz[0.7]{genbialg.tikz}
where the LHS has $m$ green nodes each of which is connected to all $n$ red nodes
and, in addition, all green and red nodes have exactly one input/output wire.
The inputs are the open-ended wires at the bottom and the outputs are the open-ended wires at the top of the diagram.
We will call this family of string diagrams $sK_{m,n}$. The RHS family we call $sS_{m,n}$
and it (necessarily) has $m$ inputs and $n$ outputs.

In order to represent this rule using our framework, we first have to describe a B-ESG grammar
that describes the LHS of (1). A good choice is the grammar $B_L=(G_L,T)$ given by:
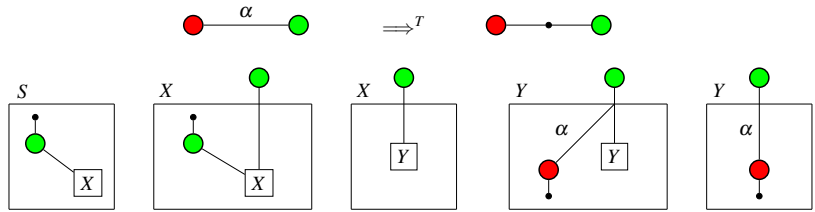
\cstikz[0.7]{bialgebra-left.tikz}
Next, we have to represent the RHS of (1). We remark that our choice of grammar has to
be consistent with the one just constructed in order to satisfy the requirements of our framework. A
suitable choice is the grammar $B_R=(G_R,T)$ given by:
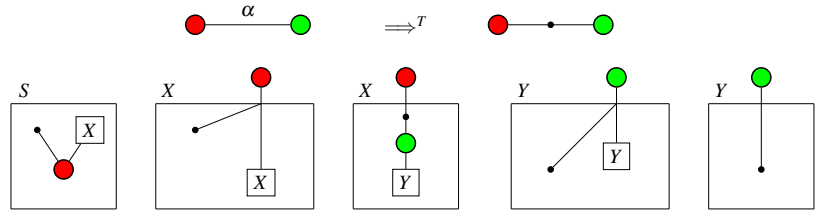
\cstikz[0.7]{bialgebra-right.tikz}
Our framework ensures that we may now uniquely construct an interface grammar by just taking the nonterminal vertices
and inputs/outputs from each production of either grammar $B_L$ or $B_R.$ We name the resulting grammar $B_I=(G_I,T):$
\cstikz[0.7]{bialgebra-middle.tikz}
There are now obvious grammar embeddings $l$ and $r$, such that $B:= (B_L \xleftarrow{l} B_I \xrightarrow{r} B_R)$ is a
B-ESG rewrite pattern. The pattern $B$ therefore encodes the rewrite rule $sK_{m,n} \leadsto sS_{m,n}.$ Of course,
by swapping $B_L$ and $B_R$ we may represent the other direction of equation (1), but we shall
only consider the left-to-right direction in this example.

Our framework guarantees that for any specific choice of $m$ and $n$, we get a string graph DPO rewrite rule $sK_{m.n} \leadsto sS_{m,n}.$
The parallel derivation of $B$ which produces the DPO rewrite rule $sK_{3,2} \leadsto sS_{3,2}$ is given by:
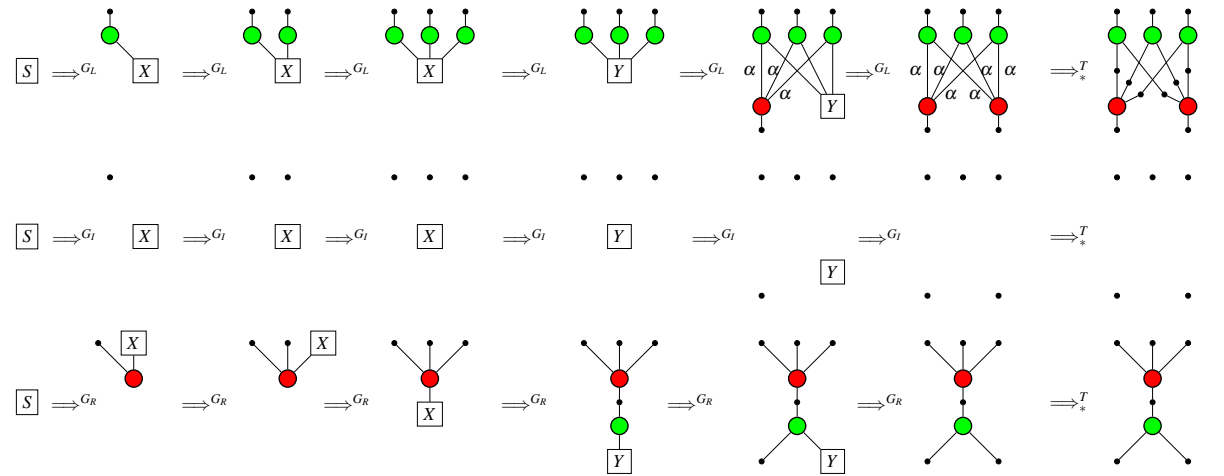
\cstikz[0.63]{32-derive.tikz}
Observe that the middle derivation is uninteresting, because it is uniquely determined by either of the other two derivations
by just taking the nonterminal vertices and the inputs/outputs in corresponding productions. Once the derivation process is over,
the result is the DPO rewrite rule $sK_{3,2} \leadsto sS_{3,2}$:
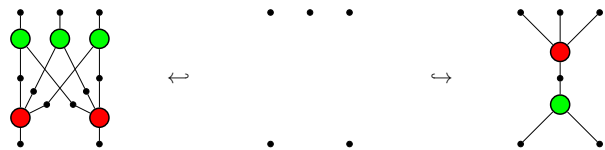
\cstikz[0.7]{32-dpo.tikz}
where the obvious embeddings are induced by the derivation process of the grammar. Again, observe that the interface of the DPO rewrite rule is
unqiuely determined by either the left or right graph by taking the inputs/outputs.

Next, we illustrate how our framework supports rewriting of families of string diagrams. Consider the family of string diagrams shown below:
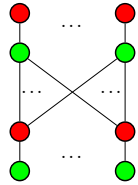
\cstikz[0.7]{family-rewrite-left.tikz}
If we apply the equational schema (1) to it, then we get the derived equational schema (2):
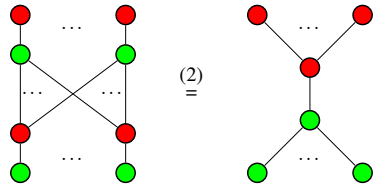
\cstikz[0.7]{family-rewrite-example.tikz}
To represent this rewrite in our framework, we start by constructing a B-ESG grammar for the LHS of (2).
We choose the grammar $B_H=(G_H,T)$ as follows:
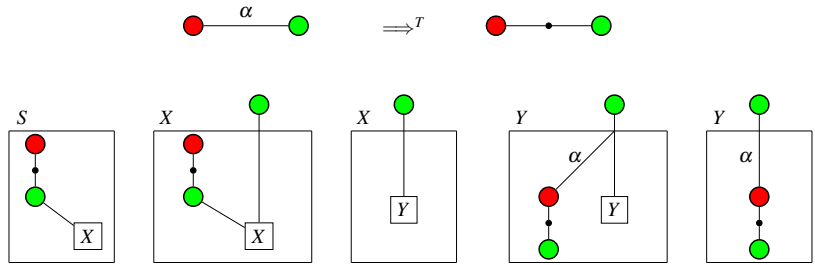
\cstikz[0.7]{big-grammar.tikz}
The previously constructed \emph{grammar} rewrite rule $B$ may now be matched into $B_H$ and the DPO rewrite can be performed, yielding the grammar $B_P=(G_P,T):$
\cstikz[0.7]{big-final.tikz}
Observe that $B_P$ correctly represents the RHS of (2) and that the \emph{derived} rewrite pattern $B' := (B_H \hookleftarrow B_I \hookrightarrow B_P)$ correctly represents the \emph{derived} equational schema (2).
More concretely, because the match satisfies the strong requirements imposed by our framework, the performed rewrite is admissible with respect to $B$ in the following sense. For any parallel derivation $(B_H,B_P) \Longrightarrow_* (H,P)$,
the string graph $H$ can be rewritten to the string graph $P$ using a sequence
of DPO rewrite rules induced by $B$ (in this specific case, this may be done
using a single DPO rule). For instance, consider the parallel derivation of $(B_H,B_P)
\Longrightarrow_* (H_{3,2}, P_{3,2})$:
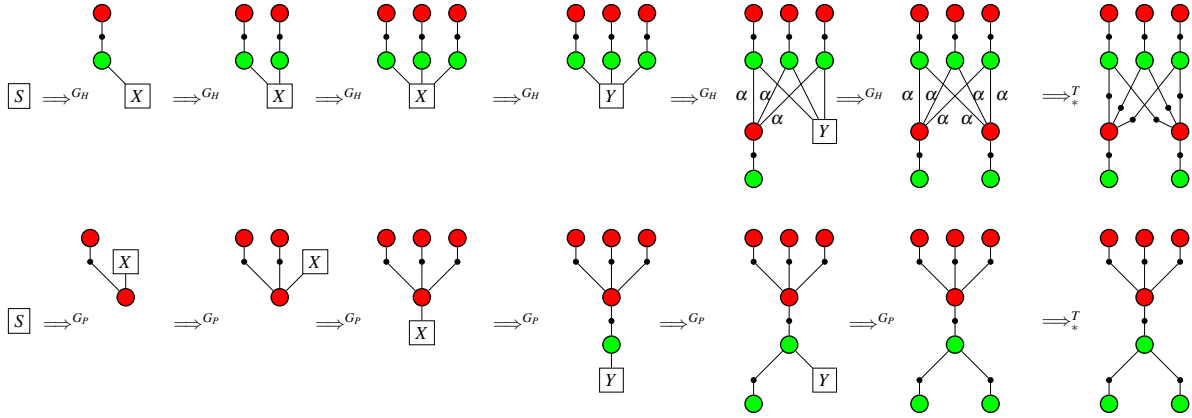
\cstikz[0.63]{final-derive.tikz}
Then string graph rewrite rule $sK_{3,2} \leadsto sS_{3,2}$ from above (which
is induced by $B$) can be matched into the string graph $H_{3,2}$ and the DPO
rewrite yields the string graph $P_{3,2}$. In this way, our framework soundly models equational reasoning with families of string diagrams.

\paragraph{Related work.}
B-ESG rewrite patterns are similar to the \textit{pair grammars}
approach presented in \cite{pair_grammars}. In that paper the author defines
a pair of graph grammars whose productions are in bijection which moreover
preserves the nonterminals within them. As a result, parallel derivations are
defined in a similar way to our B-ESG rewrite patterns. However, the author
uses a different notion of grammar which is less expressive than ours.

The pair grammars approach has inspired the development of \emph{triple
graph grammars} \cite{triple_grammars}. In this approach, the author uses a
triple of grammars $(L, C, R),$ which also share a bijective correspondence
between their productions. In this sense, they are similar to our B-ESG rewrite
rules. However, the middle grammar $C$ is used to relate graph elements from
$L$ to graph elements of $R$ in a more powerful way compared to our approach.
We simply use the middle grammar in order to identify the interface and
interior elements for performing DPO rewrites. However, the grammar model used
in \cite{triple_grammars} is based on monotonic single-pushout (SPO)
productions with no notion of nonterminal elements. These grammars are not
expressive enough for our purposes.

Another way of formalising families of string diagrams is by using
\emph{!-graphs}~\cite{pattern_graphs, merry_dphil}. !-graphs have a
considerably simpler graphical presentation compared to B-ESG grammars. This is
the underlying mechanism which Quantomatic uses for representing families of
string diagrams. Unfortunately, !-graphs also have somewhat limited expressive
power. In fact, the original motivation for developing B-ESG grammars was to
address these shortcomings. Detailed comparisons in terms of expressivity of
the two formalisms are available in~\cite{vladimir-dphil, gam}.

A third way of representing families of string diagrams is by using a
programming language designed to generate such diagrams. Examples include
Proto-Quipper-M~\cite{proto-quipper-m,eclnl} and EWire~\cite{ewire}.  Even
though these languages have not been studied in terms of their formal
expressive power, it seems very likely that they both have higher expressive
power, but worse decidability properties compared to !-graphs and B-ESG grammars
(e.g. the membership problem is unlikely to be decidable).

\section{Conclusion and future work}
We introduced B-ESG grammars which are slightly extended context-free graph
grammars that generate string graphs and which therefore represent families of
string diagrams. We showed that by carefully relating the productions of a
triple of grammars we are able to correctly represent equational schemas of
string diagrams. The category of B-ESG grammars enjoys a partial adhesive
structure and DPO rewriting in that category is not only well-behaved, but also
admissible with respect to the derivation process of the grammars, provided
that strong matching conditions are satisfied, which shows that our framework
soundly models equational reasoning about families of string diagrams.

Our framework represents string diagrams as string \emph{graphs}. More
recently, a new representation of string diagrams using \emph{hypergraphs} has
been proposed~\cite{rewriting-smt} which has a simpler and more elegant
meta-theory (e.g. no need to quotient wire-vertices on wires) and which also
enjoys better categorical properties (e.g. adhesivity vs partial adhesivity).
As part of future work, we will consider developing a new framework which can
represent families of string diagrams using \emph{hypergraph grammars}, which
also enjoy better structural and algebraic properties compared to B-edNCE
grammars. The crucial ideas presented in this paper should carry through
straightforwardly to the hypergraph setting, but it remains to be seen whether
the hypergraph representation has adequate expressive power in terms of the
languages it can generate.

\paragraph{Acknowledgements.} I would like to thank Aleks Kissinger for
many fruitful discussions and also for his supervision of my PhD studies, during which time
these results were discovered. I also thank my other former supervisors, Samson Abramsky and Bob Coecke, for
providing excellent advice during this time.

\bibliography{refs}
\end{document}

%% file: preamble.tex
\usepackage{microtype}
\usepackage{mathtools}
\usepackage{amsthm}
\usepackage{amssymb}
\usepackage{amsfonts}
\usepackage{hyperref}
\usepackage[T1]{fontenc}
\usepackage[utf8]{inputenc}
\usepackage{braket}
\usepackage{float}
\usepackage{stmaryrd}
\usepackage{wrapfig}
\usepackage{csquotes}

\usepackage[svgnames]{xcolor}
\usepackage{tikz}

\theoremstyle{plain}
\newtheorem{theorem}{Theorem}
\newtheorem{lemma}[theorem]{Lemma}

\theoremstyle{definition}
\newtheorem{definition}[theorem]{Definition}
\newtheorem{example}[theorem]{Example}

\theoremstyle{remark}
\newtheorem{remark}[theorem]{Remark}

\usetikzlibrary{decorations.pathmorphing}
\usetikzlibrary{decorations.markings}
\usetikzlibrary{decorations.pathreplacing}
\usetikzlibrary{arrows}
\usetikzlibrary{shapes}

\pgfdeclarelayer{edgelayer}
\pgfdeclarelayer{nodelayer}
\pgfsetlayers{edgelayer,nodelayer,main}

\tikzstyle{braceedge}=[decorate,decoration={brace,amplitude=10pt}]
\tikzstyle{square box}=[rectangle,fill=white,draw=black,minimum height=6mm,minimum width=6mm,yshift=0.7mm]
\tikzstyle{wire label}=[font=\footnotesize, auto,swap]

\tikzstyle{none}=[inner sep=0pt]
\tikzstyle{gn}=[circle,fill=Lime,draw=Black,line width=0.8 pt]
\tikzstyle{rn}=[circle,fill=Red,draw=Black, line width=0.8 pt]
\tikzstyle{H}=[rectangle,fill=Yellow,draw=Black]
\tikzstyle{line}=[scalar,fill=White,draw=Black]
\tikzstyle{io}=[rectangle,fill=White,draw=Black]
\tikzstyle{block}=[rectangle,fill=Orange,draw=Black]
\tikzstyle{graph}=[circle,fill=White,draw=Black]
\tikzstyle{empty}=[rectangle,fill=none,draw=none]
\tikzstyle{box}=[rectangle,fill=White,draw=Black]
\tikzstyle{dot}=[circle,fill=Black,draw=Black,inner sep=0pt,minimum size=1pt]
\tikzstyle{small dot}=[circle,fill=Black,draw=Black,inner sep=0pt,minimum size=1pt]
\tikzstyle{Dot}=[circle,fill=Black,draw=Black,inner sep=0pt,minimum size=3pt]
\tikzstyle{diam}=[rectangle,fill=Black,draw,yscale=1.2,rotate=45]
\tikzstyle{gangle}=[rectangle,fill=Lime,draw=Black]
\tikzstyle{rangle}=[rectangle,fill=Red,draw=Black]
\tikzstyle{circ}=[circle,fill=none,draw=Black,scale=1.3]
\tikzstyle{ellip}=[ellipse,fill=none,draw=Black,scale=1.3,minimum width =1.3cm]
\tikzstyle{ellip2}=[ellipse,fill=White,draw=Black,scale=1.3,minimum width =3cm]
\tikzstyle{bbox}=[rectangle,fill=Blue,draw=Blue,scale=0.6]
\tikzstyle{gg}=[shape=rectangle,fill=White,draw=Black,dashed]

\tikzstyle{nodev}=[circle,fill=none,draw=Black,scale=1]
\tikzstyle{greynode}=[circle,fill=Grey,draw=Black,scale=1]
\tikzstyle{blacknode}=[circle,fill=Black,draw=Black,scale=1]
\tikzstyle{wirev}=[circle,fill=Black,draw=Black,inner sep=0pt,minimum size=3pt]
\tikzstyle{wirevred}=[circle,fill=Red,draw=Black,inner sep=0pt,minimum size=3pt]

\tikzstyle{simple}=[-,draw=Black]
\tikzstyle{directed}=[->,draw=Black]
\tikzstyle{bdirected}=[<->,draw=Black]
\tikzstyle{bothdirs}=[bdirected,draw=Black]
\tikzstyle{bothdirsred}=[bdirected,draw=Red]
\tikzstyle{blue}=[-,draw=Blue]
\tikzstyle{redd}=[directed,draw=Red]
\tikzstyle{redu}=[-,draw=Red]
\tikzstyle{blued}=[directed,draw=Blue]
\tikzstyle{dash}=[dashed,draw=Black]
\tikzstyle{dashedred}=[dashed,draw=Red]

\tikzstyle{dotpic}=[scale=0.5]

\tikzstyle{every picture}=[baseline=-0.25em]

\newcommand{\stikz}[2][1]{\scalebox{#1}{
\InputIfFileExists{#2}{}{\input{./tikz/#2}}
}}
\newcommand{\cstikz}[2][1]{\begin{center}\stikz[#1]{#2}\end{center}}